\documentclass[12pt]{article}
\usepackage[english,activeacute]{babel}
\begin{document}
\title{Equations of motion of a relativistic charged particle with curvature
dependent actions.}
\author{Guillermo Arreaga-Garc\'\i a\footnote{garreaga$@$cifus.uson.mx}\hspace{0.25 cm} and
Julio Saucedo Morales\\
Departamento de Investigaci\'on en F\'{\i}sica de la Universidad de Sonora.\\
Apdo. Postal 14740, CP. 83000, Hermosillo, Sonora, MEXICO.}
\maketitle
\begin{abstract}
We present an introduction to the study of a relativistic
particle moving under the influence of its own Frenet-Serret
curvatures. With the aim of introducing the notation and conventions
used in this paper, we first recall the action of a relativistic
particle. We then suggest a mathematical generalization of this
action in the sense that now the action may include terms of the
curvatures of the world line generated by the particle in Minkowski
space-time. We go on to develop a pedagogical introduction to a
variational calculus which takes advantage of the Frenet-Serret
equations for the relativistic particle. Finally, we consider a
relativistic particle coupled to an electromagnetic field that is
moving under the influence of its own Frenet-Serret curvatures.
Within this frame based on the Frenet-Serret basis,
we obtain the equations of motion for several curvature dependent actions
of interest in physics. Later, as an illustration of the formalism developed, we
consider the simplest case, that of a relativistic particle when no geometrical
action is included, in order to show (i) the equivalence of this formalism
to the Newton's second law with the Lorentz force and (ii) the integrability in
the case of a constant electromagnetic field.
\end{abstract}
\newpage
\section{Introduction}
\label{int}

The dynamics of a mechanical system is governed by
{\it the principle of least action}, which states
that the motion of a system, between the times $t_i$ and $t_f$, is such
that the action $A$, defined by the integral

\begin{equation}
\label{acciongeneral}
A= \int_{t_i}^{t_f} d t \; L\, ,
\end{equation}
\noindent takes the least possible value. In general, the Lagrangian
$L$ depends only on
the positions and velocities of the system,
see Ref.\cite{landau}. Although sometimes, the motion of the
mechanical system can only be described by means of
empirical Lagrangians.

For a free relativistic particle, the lagrangian and action are given by

\begin{equation}
\label{acciongeneralfreepart}
A= -mc^2  \, \int_{t_i}^{t_f} d \, t \, \sqrt{1-v^2/c^2}
\end{equation}
\noindent where $m$ and $v$ are the mass and velocity of the
particle while $c$ is the speed of light in empty space. The
dynamics of a relativistic particle can be better studied
in {\it Minkowski space-time}, here denoted by
${\cal M}$. The points in ${\cal M}$ are called
{\it events}, which are generated by the point particle motion. An event
has three spatial coordinates $x,y,z$ and the time $t$,
The collection of events forms the
particle {\it world line} in ${\cal M}$. The length or {\it the
infinitesimal interval} of the world line on ${\cal M}$ is given by
$ds^2=-dt^2+d\vec{x}^2$. Then, following the Wienberg's conventions
in Ref.\cite{wein}, it can be written as $ds^2=\eta_{\mu \, \nu} \, dx^\mu
\; dx^\nu $, where

\begin{equation}
\eta_{\mu \, \nu}=\left\{
\begin{array}{ll}
-1 & \mu=\nu=0 \vspace{0.1 cm}\\
1  & \mu,\nu=>i,j=1,2,3 \vspace{0.1 cm}\\
0  & \mu \neq \nu
\end{array} \right.\,.
\end{equation} where $\mu=0,1,2,3$, such that $x^0=t$ and $x^i$ are
$x,y,z$. Let us now denote the particle world line by $X^{\mu}$.
The $X^{\mu}$ are the embedding functions of the particle motion
in ${\cal M}$ by means of $x^\mu=X^{\mu}(\xi)$, where $\xi$ is an arbitrary
parameter only useful for generating the motion. For example, the
{\it proper time} of the particle is given by

\begin{equation}
\label{tau2}
ds^2=-d \tau^2 = \eta_{\mu \nu} \, \frac{d X^\mu}{d \xi}
\frac{d X^\nu}{d \xi} d \xi ^2.
\end{equation}
\noindent It will be convenient
to introduce the scalar function $\gamma$ by

\begin{equation}
\label{defgamma}
\gamma= \eta_{\mu \nu} \, \frac{d X^\mu}{d \xi} \frac{d X^\nu}{d \xi}
\end{equation}
\noindent then the particle proper time is given by

\begin{equation}
\label{tau}
d \tau=\sqrt{-\gamma}\, d \xi
\end{equation}
\noindent and therefore the action of a free relativistic particle can be
written as

\begin{equation}
\label{acciongeneralfreepartgama}
A= \int_{t_i}^{t_f} \frac{d \, t}{ \gamma} \equiv \int d\, \tau
\end{equation}

It is possible to consider a mathematical generalization of the
action given in Eq.~\ref{acciongeneralfreepartgama}. The main idea
behind this generalization has come from string theory, where the string
action included an additional term proportional to the external
curvature of the string world-sheet. This additional term was proposed in order to avoid
the occurrence of sharp string configurations in the resulting string dynamics,
see Ref.~\cite{polyakov}. For this reason, it was named
the rigid string theory. The natural translation of this idea to
the action of a point particle, was to include an additional term
to Eq.\ref{acciongeneralfreepartgama} with the world-line curvature. 

In the purely mathematical sense, the relativistic particle
Lagrangian can be generalized by including terms depending on all the curvatures of the
particle world-line in the following manner. As the world-line of a
particle evolving in a fixed Minkowski space-time of general
dimension $N+1$, can have associated up to $N$ curvatures,
$\kappa_{1..N}$, a hierarchy of Lagrangians with
successively higher curvatures $\kappa_{1..N}$ can be introduced into the
action in the form

\begin{equation}
\label{acciongeneralgeom}
A[X]= \int d \tau \, L_g(\kappa_1,\kappa_2,\kappa_3,...\kappa_N).
\end{equation}
\noindent In fact, in Ref.\cite{fsdyn} such mathematical generalization was considered.

The mathematical models pointed out in Eq.~\ref{acciongeneralgeom}
have also been considered as effective theories for describing the
dynamics of an object when its internal structure is not well
resolved. For instance, an effective bosonic theory used to describe a
super-symmetric particle can be put in terms of some of these
models, see Ref.~\cite{polyakovl}. Besides, the description of spinning particles has
incorporated the attractive idea of considering that the spin degrees
of freedom are encoded in the world line geometry. In these two examples,
the extrinsic curvature in the Lagrangian is expected to supply those extra degrees of freedom.
Because the curvature is proportional
to the particle's acceleration, such effective actions will then contain terms
with derivatives on $X^{\mu} (\tau) $ equal or higher than two, that is
$\frac{ d^2 X^{\mu}}{d\, \tau^2}$. 

The theoretical study of such curvature
dependent actions for the relativistic point
particle has a long history.  Let us now mention just a few examples
of the kind of curvature dependent theories that have been
considered so far.  It was started in
the late 80s, when Plyushchay studied for the first time the physics
of the linear theory in the first curvature, that
is, $L_g= - \mu + \kappa_1$, see Ref.~\cite{canonical}. Subsequently, in
Ref.~\cite{massive}, Plyushchay was able to show that there are three types of
explicit solutions for the dynamics of
such rigid particle, namely: massive, massless and tachyonic, depending on the value of the
particle invariant $M^2 \equiv P^{\mu}\, P_{\mu}$, where $P^\mu$ is particle's
four momentum.  Dereli {\it et al.} in Ref.~\cite{dereli} have considered the
theory $L_g= - \mu + \kappa_1^2$ while
Nesterenko {\it et al.} ~\cite{neste} and ~\cite{neste2} have
studied the integrability of the case $L_g= f(\kappa_1) $
for a general function $f$. The linear model in the first curvature, $L_g= \kappa_1$,
was firstly considered by Plyushchay in Ref.~\cite{massless} and by
Zoller in Ref.~\cite{zoller} who was able to show that
the natural coupling of this linear curvature theory to gravity does not yield a
consistent theory.

Plyushchay went on to consider the most
general curvature dependent lagrangian, $L_g= f(\kappa_1^2) $, with the purpose
of studying the quantization of such a particle, see Ref.~\cite{quanti}.
The question of quantization of a relativistic particle models with higher
derivatives was investigated again by Plyushchay in Ref.~\cite{otro}.

The model of the relativistic
particle with an action depending linearly on the curvature and torsion,
$L_g= - \mu + cte \, \kappa_1 + cte \,\kappa_2$, was investigated both at the
classical and quantum levels by Kuznetsov
and Plyushchay, see Refs.~\cite{kuzne} , \cite{comment}, \cite{curvyt1} and \cite{curvyt2}.
A review of the relations between the Majorana equation to the
higher derivative particle models was provided in Ref.\cite{majorana}.  Particularly,
the model with torsion has been proved to be usefull as a model for relativistic anyons in
$(2+1)D$ space-time, see Refs.~\cite{partorsion1} and \cite{partorsion2}. A relation between
the relativistic particle with torsion in Minkowski $(2+1)D$ spacetime to the model of the
non-relativistic $3D$ Euclidean particle in the field of a monopole has been
revealed in Ref.~\cite{mono}.  Plyushchay also demonstrated that the equation of
motion of a particle including a linear term
on the second curvature ( or torsion) of the world-line, coincide with the
equation of motion of a relativistic charged particle in an external constant
electromagnetic field, see Ref.~\cite{iden}.

More recently in Ref.\cite{barros}, Barros has considered a Lagrangian depending linearly
on the world line curvature with the purpose of describing the
dynamics associated with relativistic particles both massive and
massless. He has obtained the whole space of solutions in an
space-time with constant curvature. In
Ref.~\cite{angel}, Ferr\'andez {\it et al.} have shown the moduli spaces of
solutions in a three dimensional pseudo-Riemannian space for the model of a relativistic particle
with curvature and torsion. In Ref.~\cite{josu}, Arroyo {\it et al.} have considered the model
$L_g= m + n \, \kappa_1 + p \,\kappa_2$ and has found
all the solutions for the constants $m,n,p$; or in other words, that the spinning
relativistic particles evolve along Lancret curves in a $3D$ space with constant curvature. In
Ref.~\cite{ferra} numerical integration of the Cartan equation of motion for a relativistic
particle with curvature was carried out in order to explicitly show the particle world-lines.

The authors of Refs.\cite{lopez} and \cite{fomiga}, have explored the
correspondence between the geometry of the world lines described by
a Frenet-Serret basis and the dynamics of a charged particle as they
related the two invariants of the electromagnetic field with the
curvatures of the world line.

\section{Geometric elements}
\label{sec:geomele}

The mathematical problem of finding the equation of motion
associated with an action implies the calculation of the variation
of all the terms under the action integral, in such a way that the
real motion of the particle will be the one having its first
variation identically zero.

A Frenet-Serret vectorial basis can be attached to every point of
any smooth curve in $3D$ space generated by the motion of a
particle. The basis consists of a set of three vectors: the tangent
vector and two ortho-normal vectors, which are called the normal and
the bi-normal, respectively. To complete the Frenet-Serret basis, one also
needs to introduce the scalar functions $\kappa$ and $\tau$,
known as the curvature and torsion, respectively, see Ref.
\cite{eisen}.

Thus, in order to calculate the first variation of all the geometrical terms
involved in an action like the one given
by Eq.~\ref{acciongeneralgeom}, we
develop in Section~\ref{subsec:pertur} a variational
procedure, based on the
Frenet-Serret equation for a relativistic particle. But we first
show in Section ~\ref{subsec:tFSb} the generalization of a well
know procedure in $3D$ space for constructing the Frenet-Serret equations
in the four dimensional spacetime ${\cal M}$.

The Frenet-Serret basis adapted for the Minkowski spacetime ${\cal M}$
has been used to generalize the fundamental
theorem of curves in Euclidean $3D$ space, so that the curvature and
torsion completely determine the curve up to a rigid motion.  In
Ref.~\cite{fomiga} was shown that in the context of
Minkowski spacetime, the three curvatures fixed the world line
up to a Poincare transformation.

The authors of \cite{fomiga} also offered a proof for the statement
that a world line with a non-vanishing first curvature is plane if
and only if the second and third curvatures identically vanish. For
the case in which just the third curvature vanishes, the world
line lies in a hyperplane.

\subsection{The Frenet-Serret basis}
\label{subsec:tFSb}

Let us now generalize the idea of introducing a vectorial basis at any
given point of a curve generated by a relativistic particle
evolving in space-time ${\cal M}$. Let us start by defining
the tangent vector $T^\mu$ by
$T^\mu=\frac{d \, X^{\mu} }{d \, \tau} \equiv X'^\mu$, where a
prime hereafter means a derivative with respect
to $\tau$, the proper time defined by Eq.\ref{tau}.
Making the chance of coordinates
$\xi \, => \, \tau$ in Eq. \ref{tau2}, we get

\begin{equation}
\label{ortot}
\eta(T,T)\equiv \eta_{\mu \nu} T^\mu T^\nu =-1\,.
\end{equation}
\noindent The set of four-vectors $N_i^\mu$ must be orthogonal to
the tangent four-vector $T^{\mu}$, as well as unitary. By taking the derivative of
Eq.~\ref{ortot} with respect to $\tau$, we find that $\eta(\frac{d
T}{d\tau},T)=0$, and that these vectors are orthogonal. We then
define the four-vector $N_1^{\mu}$ by

\begin{equation}
\label{fs1}
\frac{d T^\mu}{d \tau} \equiv \kappa_1 \, N_1^\mu
\end{equation}
\noindent such that $\kappa_1$ is its norm, and it satisfies
the relations

\begin{equation}
\label{ortovectores1}
\eta(T,N_1)=0 \hspace{0.5 cm}  \eta(N_1,N_1)=1
\end{equation}
\noindent
\noindent Let us now take the derivative $\frac{d }{d \tau}$ of the
first relation of Eq.~\ref{ortovectores1}:
$\eta(\frac{d T}{d\tau},N_1)+ \eta(T,\frac{ d N_1}{d \tau})=0$. By using
Eq.~\ref{fs1} into the left hand side of this last expression, we have that
$\kappa_1 + \eta_{\mu \nu} T^\mu \frac{ d N^\nu_1}{d \tau}$. It is possible
to factorize the tensor $\eta_{\mu \nu}$ from this equation, to obtain
$\eta_{\mu \nu} T^\mu \left( -\kappa_1 T^\nu + \frac{ d N^\nu_1}{d \tau}\right)=0$
from which we conclude that the vector in the bracket is orthogonal to the
tangent, therefore we can define the second normal vector by

\begin{equation}
\label{fs2}
\frac{d N_1^\mu}{d \tau} \equiv \kappa_2 \, N_2^\mu + \kappa_1 \, T^\mu
\end{equation}
\noindent such that $\kappa_2$ is its norm; then the four-vector $N_2^{\mu}$ satisfies
the relations

\begin{equation}
\label{ortovectores2}
\eta(T,N_2)=0 \hspace{0.5 cm} \eta(N_2,N_2)=1 \hspace{0.5 cm} \eta(N_1,N_2)=0
\end{equation}
\noindent That $N_2$ is also orthogonal to $N_1$ can easily be proved
by dotting Eq.~\ref{fs2} with $N_1$.

Let us now continue by taking the derivative of the third
relation of Eq.~\ref{ortovectores2} with respect to $\tau$. We have
$\eta(\frac{ d N_1}{d \tau} ,N_2) + \eta(N_1,\frac{ d N_2}{d
\tau})=0$. Substituting Eq.~\ref{fs2} into the first term of this
relation and factorizing the term $\eta(N_1,)$ in its components
explicitly, we obtain $\eta_{\mu \nu} N_1^\mu \left( \kappa_2
N_1^\nu + \frac{ d N^\nu_2}{d \tau}\right)=0$, from which we are
able to introduce the definition of the third normal four-vector $N_3^{\mu}$ by
means of

\begin{equation}
\label{fs3}
\frac{d N_2^\mu}{d \tau} + \kappa_2 \, N_1^\mu \equiv \kappa_3 \; N^\mu_3
\end{equation}
\noindent such that $\kappa_3$ is its norm; then the vector $N_3$ satisfies
the relations

\begin{equation}
\label{ortovectores3}
\eta(T,N_3)=0 \hspace{0.5 cm}  \eta(N_3,N_3)=1 \hspace{0.5 cm} \eta(N_1,N_3)=0  \hspace{0.5 cm} \eta(N_2,N_3)=0 \\
\end{equation}
\noindent By dotting Eq.~\ref{fs3} with $N_2$ we obtain the third
relation of Eq.~\ref{ortovectores3}. Likewise, by dotting Eq.~\ref{fs3}
with $T^\mu$ and commuting the derivative between the terms and
substituting Eq.~\ref{fs1}, we prove the first relation of
Eq.~\ref{ortovectores3}.

Finally, we take the derivative with respect to $\tau$ of the third
relation of Eq.~\ref{ortovectores3}: $\eta(\frac{ d N_2}{d \tau}
,N_3) + \eta(N_2,\frac{ d N_3}{d \tau})=0$. Substituting
Eq.~\ref{fs2} into the first term of the precedent expression and
factorizing the term $\eta(N_2,)$, we get $\eta_{\mu \nu} N_2^\mu
\left( \kappa_3 N_2^\nu + \frac{ d N^\nu_3}{d \tau}\right)=0$. We
could introduce a fourth normal vector defining it as the term
inside the parenthesis. However, there is no dimension in the spacetime
${\cal M}$ to associate another normal vector, so we have to
close the process and define the term as identically zero, therefore

\begin{equation}
\label{fs4}
\frac{d N_3^\mu}{d \tau} + \kappa_3 \, N_2^\mu \equiv 0
\end{equation}

Summing up, so far we have built the Frenet-Serret basis
for the relativistic particle in the space ${\cal M}$; a
basis which is formed by the following vectors

\begin{equation}
\label{eq:fs}
\begin{array}{ll}
\frac{d T^\mu}{d \tau}& =\kappa_1 \, N_1^\mu
\vspace{0.2 cm}\\
\frac{d N_1^\mu}{d \tau}& = \kappa_2 \, N_2^\mu + \kappa_1 \, T^\mu
\vspace{0.2 cm}\\ \frac{d N_2^\mu}{d \tau}& = \kappa_3 N_3^\mu-\kappa_2 \, N_1^\mu
\vspace{0.2 cm}\\
\frac{d N_3^\mu}{d \tau}& = -\kappa_3 \, N_2^\mu \\
\end{array}
\end{equation}
\noindent and with the following properties

\begin{equation}
\label{ortovectores}
\begin{array}{l}
\eta(T,N_i)=0 \hspace{0.2 cm} i=1,2,3\\
\eta(N_i,N_j)=\delta_{ij}=\left\{
\begin{array}{l}
0 \hspace{0.2 cm} i \neq j \\
1 \hspace{0.2 cm} i=j
\end{array}\right.
\end{array}
\end{equation}
\noindent The system of equations ~\ref{eq:fs} can be
easily generalized for a curved space-time by making on all the terms
the transformations of derivatives $\frac{d A^\mu}{d \tau}$ to
covariant derivatives, that is, $\frac{D A^\mu}{D \tau } =\frac{d A^\mu}{d \tau }
+ \Lambda^{\mu}_{\alpha \beta} \frac{d X^\alpha}{d \tau } A^{\beta}$.

\subsection{Relativistic Perturbations.}
\label{subsec:pertur}

Let us now consider a small perturbation on a general
particle world line embedded in the
spacetime ${\cal M}$ by the functions $x^\mu=X^\mu(\tau)$. Let us define
$\delta $ as a perturbation operator, which action on $X^\mu(\tau)$ can be decomposed
along the Frenet-Serret vectors as

\begin{equation}
\label{pertur}
\delta X^\mu= \psi_{||} T^\mu + \psi_1 N_1^\mu + \psi_2 N_2^\mu
+ \psi_3 N_3 ^\mu
\end{equation}
\noindent where the functions $\psi_{||},\psi_i$ are
numerically small in such a way that we only consider
the linear perturbation regime.

By taking the derivative of any vector with respect to $\tau$, we get
another vector, which can also be expressed as a linear combination
of the Frenet-Serret vectors, for example

\begin{equation}
\label{comderdeltaX}
\frac{d\, \delta X^\mu}{d \tau }=
\left( \psi_{||}' + \kappa_1 \psi_i \right) T^\mu +
\left( \psi_1' - \kappa_2 \psi_2 + \kappa_1 \psi_{||} \right) N_1^\mu +
\left( \psi_2' - \kappa_3 \psi_3 + \kappa_2 \psi_1 \right) N_2^\mu +
\left( \psi_3' + \kappa_3 \psi_2 \right) N_3^\mu .
\end{equation}

Let us now consider the perturbation on the function $\gamma$ defined
in Eq~\ref{defgamma}. We apply
to it the variation operator $\delta$, we get

\begin{equation}
\label{vargamma}
\begin{array}{ll}
\delta  \gamma & = 2 \eta_{\mu \nu} \, \frac{d X^\mu}{d\xi } \,
\frac{d \delta X^\nu}{d \xi} \\
&= 2 \eta_{\mu \nu} \, T^\mu  \,
\frac{d \delta X^\nu}{d \tau} \, \left( \frac{d \tau }{d \xi}\right)^2\\
&= 2 \left( \frac{d \tau }{d \xi}\right)^2
\eta(T,\frac{d \delta X}{d \tau})
\end{array}
\end{equation}
\noindent where we have used the chain rule in both terms of the
dot product shown in
the second line, in order to switch to derivatives on $\tau$
instead of $\xi$. Now,
according to the third line of Eq. ~\ref{vargamma},  we only need to take
the tangential component of Eq.\ref{comderdeltaX} to obtain

\begin{equation}
\label{var:gamma}
\delta \gamma = 2 \gamma \left(\psi_{||}' + \kappa_1 \, \psi_1 \ \right)\;,
\end{equation}
\noindent the perturbation
of the proper time. Then, by applying the
operator $\delta$ on Eq.~\ref{tau}
and considering that there is a commutation relation with the derivative
operator with respect to the arbitrary parameter $\xi$

\begin{equation}
\label{conmurel}
\delta \left( \frac{d \tau}{d \xi } \right) = \frac{ d \left( \delta \tau \right)}{d \xi}
\end{equation}
\noindent we get

\begin{equation}
\label{var:tau}
\delta \left( d \, \tau \right) =  \left( d \, \tau \right)
\left(\psi_{||}' + \kappa_1 \, \psi_1 \ \right)\,.
\end{equation}
\noindent It is important to point out that the result of Eq.\ref{var:tau}
indicates that
the perturbation operator and the derivative operator with respect to the proper time
do not commute. For this reason, the chain rule must be used when there are derivatives
with respect to $\tau$. As an example, let us consider the perturbation on a derivative
of a scalar function  $f(\tau)$,

\begin{equation}
\label{var:derf}
\begin{array}{ll}
\delta \left( \frac{d  f}{d \, \tau } \right) & =
\delta \left( \frac{d  f}{d  \xi} \right) \frac{d  \xi}{d \tau}
+ \frac{d  f}{d  \tau} \delta \left( \frac{ d  \xi}{d  \tau}
\right)\\
&= \frac{d  \delta f}{d  \xi} \frac{d  \xi}{d  \tau}
+ \frac{d  f}{d  \tau} \delta \left( \frac{ d  \xi}{d  \tau}
\right)
\end{array}
\end{equation}
\noindent and replacing Eqs.\ref{tau} and \ref{var:gamma}, we get

\begin{equation}
\label{var:derft}
\delta \left( \frac{d \, f}{d \, \tau } \right) =
\frac{d \left(\delta f \right) }{d \tau }-\frac{d f}{d \tau }
\, \left(\psi_{||}' + \kappa_1 \, \psi_1 \ \right) \;.
\end{equation}
\noindent We can now obtain the perturbation of the tangent
four-vector by considering the change $f=>X^\mu$ in Eq.~\ref{var:derft}, we get

\begin{equation}
\label{var:tangdef}
\delta T^{\mu}  \equiv \delta \left( \frac{d X^\mu}{d\, \tau} \right) =
\frac{d \delta X^\mu}{d \tau} - T^{\mu} \left(\psi_{||}' + \kappa_1 \, \psi_1 \ \right)
\end{equation}
\noindent and comparing with Eq.\ref{comderdeltaX} we get that
the perturbation of $\delta T^\mu$

\begin{equation}
\label{var:tang}
\delta T^{\mu} =
\left( \psi_1' - \kappa_2 \psi_2 + \kappa_1 \psi_{||} \right) N_1^\mu +
\left( \psi_2' - \kappa_3 \psi_3 + \kappa_2 \psi_1 \right) N_2^\mu +
\left( \psi_3' + \kappa_3 \psi_2 \right) N_3^\mu .
\end{equation}
\noindent is purely orthogonal to the world-line. For mathematical
convenience, we now introduce the $\alpha$ coefficients in the following form

\begin{equation}
\label{var:tangalfas}
\delta T^{\mu}  = \alpha_1 \, N_1^{\mu} + \alpha_2 \, N_2^{\mu} +
\alpha_3 \, N_3^\mu
\end{equation}
\noindent where

\begin{equation}
\label{alfas}
\begin{array}{ll}
\alpha_1 & = \psi_1'- \kappa_2 \psi_2 + \kappa_1 \psi_{||} \\
\alpha_2 & = \psi_2'- \kappa_3 \, \psi_3 +\kappa_2 \psi_1 \\
\alpha_3 & = \psi_3'+\kappa_3 \, \psi_2 \,.
\end{array}
\end{equation}

With the purpose of calculating the
variation of the first curvature, we solve for $\kappa_1$ from the first
Frenet-Serret Eq.~\ref{fs1}, obtaining $\kappa_1=\eta (N_1, \frac{d\, T}{d\, \tau })$.
Then, applying the $\delta$ operator we get,

\begin{equation}
\label{var:kappa1def}
\delta \kappa_1 = \eta(\delta N_1, \frac{d\, T}{d\, \tau })
+ \eta (N_1,\delta \frac{d\, T}{d\, \tau }).
\end{equation}
\noindent The first term vanishes, as it can be probed by making use again
of the first Frenet-Serret Eq.~\ref{fs1} and factorizing the operator $\delta$, that is
$\eta(\delta N_1, \frac{d\, T}{d\, \tau })= \kappa_1 \eta(\delta N_1,N_1)
=\frac{\kappa_1}{2} \delta \eta(N_1,N_1) \equiv 0\, $. To calculate the second
term of Eq. \ref{var:kappa1def}, we can use Eq.\ref{var:derft} with $f=>T^\mu$, to get

\begin{equation}
\label{var:dTdtau}
\delta \frac{d T^\mu}{d \tau }=
\frac{d\delta  T^\mu}{d\, \tau }-
\frac{dT^\mu}{d\tau}\left(\psi_{||}' + \kappa_1 \psi_1  \right)
\end{equation}
\noindent Using now Eqs. \ref{comderdeltaX} and \ref{var:tangalfas}
but with $\alpha$ coefficients $\alpha$ defined by Eq. \ref{alfas} instead of the
$\psi$ functions, we get the component along the $N_1$ vector, in such a way that
the variation is,

\begin{equation}
\label{var:kappa1alfas}
\delta \kappa_1 = \alpha_1' -\alpha_2 \kappa_2 -
\kappa_1\left(\psi_{||}'+ \kappa_1 \psi_1 \right)
\end{equation}
\noindent So, the final result of this variation expressed in terms
of the $\psi$ functions is

\begin{equation}
\label{var:kappa1final}
\delta \kappa_1 = \psi_1'' -\left( \kappa_1^2 + \kappa_2^2 \right) \psi_1-
2  \kappa_2 \psi_2' -\kappa_2' \psi_2 + \kappa_2 \kappa_3 \psi_3 -
\kappa_1 \psi_{||}'' + \kappa_1' \psi_{||}
\end{equation}

Following the same method, we can calculate the variation to first order
for all the other curvatures.  To get $\delta \kappa_2$ we first
calculate $\delta N_1^\mu$,  by solving for
the second curvature in the second FS Eq.~\ref{fs2} and applying the perturbation
operator $\delta$, we obtain

\begin{equation}
\label{var:kappa2def}
\delta \kappa_2 = \delta \eta ( N_2,\frac{d N_1}{d \tau})=
\eta (\delta  N_2, \frac{d  N_1}{d \tau}) +
\eta (N_2, \delta \frac{d  N_1}{d \tau})
\end{equation}
\noindent Substituting the second FS Eq. \ref{fs2} in the first term
and using the commutation relation \ref{var:derft} in the second term, we get

\begin{equation}
\label{var:kappa2int1}
\delta \kappa_2 =  \kappa_1 \eta (\delta N_2 , T) +
\kappa_2 \eta (\delta N_2,N_2 ) +
\eta(N_2,\frac{d \delta  N_1}{d\, \tau})-
\left(\psi_{||}' + \kappa_1 \psi_1 \right)\eta ( N_2,\frac{d N_1}{d\tau})
\end{equation}
\noindent We notice that the second term of Eq.~\ref{var:kappa2int1}
vanishes and that we can use again the
second FS Eq.~\ref{fs2} to simplify the
fourth term.  The fact that $\eta$ is
constant and that the FS vectors are all orthogonal allow us to interchange the action
of the operator $\delta$ on the first term
$\eta(T,\delta N_2)=-\eta(\delta T, N_2)$; we also notice that the third term can be
written in the form

\begin{equation}
\label{var:kappa2int2}
\eta(N_2,\frac{d \delta N_1}{d\tau}) =
\frac{d}{d \tau} \eta(N_2,\delta N_1) -
\eta(\frac{d N_2}{d\tau},\delta N_1) \,.
\end{equation}
\noindent so that Eq. \ref{var:kappa2int2} becomes

\begin{equation}
\label{var:kappa2int2o}
\delta \kappa_2 =  - \kappa_1 \eta (N_2, \delta T ) -
\kappa_3 \eta(N_3,\delta N_1 )-
\left(\psi_{||}'' + \kappa_1 \psi_1 \right) \kappa_2
+\frac{d}{d\tau} \eta(N_2,\delta N_1)
\end{equation}
\noindent At this point we note that for calculating $\delta
\kappa_2$ one only needs to compute the components along the
directions $N_2$ and $N_3$ of the variation $\delta N_1$, as we know
that the component along the $N_2$ direction of $\delta T^\mu $ is
$\alpha_2$.

Let us now calculate the required components of $\delta N_1$. Solving for $N_1$ from
the first FS Eq.\ref{eq:fs} and applying the operator $\delta$, we have

\begin{equation}
\label{var:n1}
\delta N_1^\mu=
\delta \left( \frac{1}{\kappa_1} \right) \frac{d T^\mu}{d \tau}
+ \left( \frac{1}{\kappa_1} \right) \delta \frac{d T^\mu}{d \tau}
\end{equation}
\noindent and making use of Eq. \ref{var:derft}, this becomes

\begin{equation}
\label{var:n1term1}
\delta N_1^\mu = -\left( \frac{\delta \kappa_1}{\kappa_1} \right)
N_1^\mu + \frac{1}{\kappa_1}  \frac{d \delta T^\mu}{d \tau} -
\frac{1}{\kappa_1}
\frac{d T^\mu}{d \tau} \left( \psi_{||}' + \kappa_1 \psi_1 \right)
\end{equation}
\noindent Obviously, the needed component is contained in the second
term of Eq. \ref{var:n1term1}. By once againg using Eq.\ref{comderdeltaX} with
the $\alpha$ coefficients instead of the $\psi$ functions, we get the
second term of Eq.\ref{var:kappa2int2o}, that is,

\begin{equation}
\label{var:n1compn3}
\eta(N_3,\delta N_1)=
\frac{1}{\kappa_1} \eta(N_3,\delta \frac{d T^\mu}{d\tau})=
\left( \frac{1}{\kappa_1} \right)
\left( \alpha_3' + \alpha_2 \kappa_3 \right)
\end{equation}
\noindent we finally get the variation $\kappa_2$ in compact notation

\begin{equation}
\label{var:kappa2finalcom}
\delta \kappa_2 = \kappa_1 \alpha_2 -
\frac{\kappa_3}{\kappa_1} \left(\alpha_3' + \kappa_3 \alpha_2 \right)
- \kappa_2 \left( \psi_{||}' + \kappa_1 \psi_1 \right)
+ \frac{d}{d\tau}
\left( \frac{\alpha_2' + \alpha_1 \kappa_2 - \alpha_3 \kappa_3}{\kappa_1}\right)
\end{equation}
\noindent or in terms of the $\psi$ functions, this variation becomes

\begin{equation}
\label{var:kappa2final}
\begin{array}{ll}
\delta \kappa_2 = & -\left( \frac{\kappa_3}{\kappa_1} \right) \psi_3'' +
\left( \kappa_1 \kappa_3 + \frac{\kappa_3^3}{\kappa_1}\right) \psi_3
-\left( \kappa_1 + 2 \frac{\kappa_3^2}{\kappa_1} \right) \psi_2'
- \left( \frac{\kappa_3 \kappa_3'}{\kappa_1}\right) \psi_2
-\left( 2 \kappa_1 \kappa_2
+ \frac{\kappa_2 \kappa_3^2}{\kappa_1}\right) \psi_1 -\kappa_2 \psi'_{||}\,.
\end{array}
\end{equation}

We proceed in the same way to calculate the variation of $\kappa_3$.
We start by taking the dot product of the vector
$N_2^\mu$ with all the terms of the fourth FS Eq.~\ref{fs4}, that is,
$\kappa_3=-\eta(N_2,\frac{d N_3}{d \tau})$. Then we apply upon it
the $\delta$ operator

\begin{equation}
\label{var:kappa3ini}
\begin{array}{ll}
\delta \kappa_3 &= - \eta(\delta N_2, \frac{d N_3}{d \tau} ) -
\eta(N_2,\delta \frac{d N_3}{d\tau})\\
&= \kappa_3 \eta(\delta N_2,N_2)- \eta\left(N_2,\frac{d \delta N_3}{d\tau}
-\frac{d N_3}{d\tau}\left( \psi_{||}' + \kappa_1 \psi_1 \right) \right)\\
&= - \eta(N_2,\frac{d}{d\tau} \delta N_3 )-
\kappa_3 \left(\psi_{||}'+\kappa_1 \psi_1 \right)
\end{array}
\end{equation}
\noindent For calculating the first term of the third line, it is
better to make a double transposition of the operator, such that

\begin{equation}
\label{var:kappa3segterm}
\begin{array}{ll}
\eta(N_2,\frac{d}{d\tau} \delta N_3)&=
-\eta(\frac{d N_2}{d\tau},\delta N_3)+ \frac{d}{d\tau}\eta(N_2,\delta N_3)\\
& = \kappa_2 \eta(N_1,\delta N_3) + \frac{d}{d\tau}\eta(N_2,\delta N_3)\\
& = -\kappa_2 \eta(\delta N_1,N_3) + \frac{d}{d\tau}\eta(N_2,\delta N_3)\\
\end{array}
\end{equation}
\noindent then, in compact notation the variation is

\begin{equation}
\label{var:kappa3finalcom}
\delta \kappa_3 = \frac{\kappa_2}{\kappa_1} \left(\alpha_3'
+ \kappa_3 \alpha_2 \right)
- \kappa_3 \left( \psi_{||}' + \kappa_1 \psi_1 \right) - \frac{d}{d\tau}\eta(N_2,\delta N_3)
\end{equation}
\noindent Again, it is convenient to express this result in terms of the $\psi$ deformation
functions, then

\begin{equation}
\label{var:kappa3finalpsi}
\delta \kappa_3 = \frac{\kappa_2}{\kappa_1}
\left[
\left( \psi_3'' - \kappa_3^2 \psi_3 \right) +
\left( 2\kappa_3 \psi_2' + \kappa_3' \psi_2 \right)
\right]
+\left( \frac{\kappa_2^2 \kappa_3}{\kappa_1}- \kappa_1 \kappa_3 \right)\psi_1
-\kappa_3 \psi_{||} - \frac{d}{d\tau}\eta(N_2,\delta N_3) \,.
\end{equation}

It should be noted that Eqs. \ref{var:kappa1final} ,
\ref{var:kappa2final} and \ref{var:kappa3finalpsi}, with the
variation of the curvatures $\delta \kappa_1$, $\delta \kappa_2$ and
$\delta \kappa_3$, are the main results of this section.
Despite the fact that the total derivative
appearing in those equations does not make any contribution
to the equation of motion, these terms are important
for calculating the conservation laws, as can be
seen in Ref.~\cite{conslaws}.

\section{Geometric actions and equations of motion.}
\label{sec:acciones}

The action describing the dynamics of a relativistic
charged particle in the presence of an electromagnetic field is

\begin{equation}
\label{acciongeneralconcampo}
A_u[X]= \int d \tau \, \left[ -m - q A_\mu T^\mu \right] \,.
\end{equation}
\noindent where $A^\mu$ are the field potentials. This action
describes the motion of an electron in an electromagnetic field.
When a term of self-force due to the electron's own charge is
considered, the motion can be described by the Dirac-Lorentz
equation. It should be noticed that Rohrlich in Ref.\cite{roch} has
used the FS equations for a $3D$ space time with the purpose of
obtaining solutions of the Lorentz-Dirac equations. The term of
self-force on the electron includes the particle velocity and
certain terms involving the first and second derivatives of the
particle velocity.  The author of Ref.~\cite{intri} discovered that
mathematical structure of the external force can be obtained by
manipulating the Frenet-Serret vectors along the particle world
line.

Let us now consider a theory based on the Lagrangian introduced in
Eq. \ref{acciongeneralconcampo} plus a geometric
Lagrangian $L_g$, which may include world-line curvatures,
that is

\begin{equation}
\label{acciongeneralgeo}
A[X]= \int d \tau \,
\left[ -m + m_g L_g(\kappa_1,\kappa_2,\kappa_3) -q A_\mu T^\mu \right] \,.
\end{equation}
\noindent where $m_g$ is a parameter.  The case when $L_g$ is a quadratic function
on the first curvature $L_g=\kappa_1^2$ has been studied
widely by several authors, see Sect.\ref{int}.

Let us now consider the general variation of the action given in
Eq.~\ref{acciongeneralgeo}

\begin{equation}
\label{varacciongeneral}
\begin{array}{ll}
\delta A[X]& = \int \left[
\delta \left( d \tau \right) L + d \tau \, \delta L \right]
\vspace{0.1 cm}\\
& =
\int d \tau \left[
\left( \kappa_1 \psi_1 \right) L +
m_g \delta L_g(\kappa_1,\kappa_2,\kappa_3)
-q \delta A_\mu -q A_\mu \delta T^\mu  \right] \,.
\end{array}
\end{equation}
\noindent By making use of the general variations

\begin{equation}
\label{varAyLg}
\begin{array}{l}
\delta L_g =
\frac{\partial L_g}{\partial \kappa_1 }\delta \kappa_1+
\frac{\partial L_g}{\partial \kappa_2 }\delta \kappa_2+
\frac{\partial L_g}{\partial \kappa_3 }\delta \kappa_3
\vspace{0.1 cm} \\
\delta A_\mu= \frac{\partial A_\mu }{\partial X^\nu }\delta X^\nu
= A_{\mu , \nu}
\left( \psi_1 N_1^\nu + \psi_2 N_2^\nu + \psi_3 N_3^\nu
\right)
\end{array}
\end{equation}
\noindent we are now able to write down some special cases
of Lagrangian $L_g$.

\begin{itemize}
\item The model $L_g=\frac{1}{2} \kappa_1^2$, whose dynamical equations
take the form

\begin{equation}
\label{eqk1}
\begin{array}{l}
-m\kappa_1 - m_g
\left[
+\kappa_1''-\frac{1}{2}\kappa_1^3-\kappa_1 \kappa_2^2
\right]-
q F_{\mu \nu} T^\mu N_1^\nu \equiv 0
\vspace{0.1 cm} \\
-2m_g \kappa_1' \kappa_2-m_g \kappa_1 \kappa_2'
-q F_{\mu \nu} T^\mu N_2^\nu \equiv 0
\vspace{0.1 cm} \\
-m_g \kappa_1 \kappa_2 \kappa_3
-q F_{\mu \nu} T^\mu N_3^\nu \equiv 0
\end{array}
\end{equation}
\item The model $L_g=\frac{1}{2} \kappa_2^2$ whose equations of motion are given by

\begin{equation}
\label{eqk2}
\begin{array}{l}
-m\kappa_1 - m_g
\left[
\frac{1}{2} \kappa_1 \kappa_2^2 + \kappa_2^2
\left(2 \kappa_1  + \frac{\kappa_3^2}{\kappa_1} \right)
\right]-
q F_{\mu \nu} T^\mu N_1^\nu \equiv 0
\vspace{0.1 cm} \\
-m_g \frac{d}{d\tau} \left( \kappa_1 \kappa_2 +
\frac{2 \kappa_2 \kappa_3^2}{\kappa_1} \right)+
m_g \left( \frac{\kappa_2 \kappa_3 \kappa_3'}{\kappa_1} \right)
-q F_{\mu \nu} T^\mu N_2^\nu \equiv 0
\vspace{0.1 cm} \\
-m_g \kappa_1 \kappa_2 \kappa_3 - m_g
\frac{\kappa_2 \kappa_3^2}{\kappa_1}
-q F_{\mu \nu} T^\mu N_3^\nu \equiv 0
\end{array}
\end{equation}
\item The model $L_g=\frac{1}{2} \kappa_3^2$ whose equations of motions are

\begin{equation}
\label{eqk3}
\begin{array}{l}
-m\kappa_1 - m_g
\left[
\frac{1}{2} \kappa_1 \kappa_3^2 + \kappa_3^2
\left(\kappa_1 - \frac{\kappa_2^2}{\kappa_1} \right)
\right]-
q F_{\mu \nu} T^\mu N_1^\nu \equiv 0
\vspace{0.1 cm} \\
2m_g
\frac{d }{d\tau} \left( \frac{\kappa_2 \kappa_3^2}{\kappa_1} \right)
-m_g \frac{\kappa_2\kappa_3 \kappa_3'}{\kappa_1}
-q F_{\mu \nu} T^\mu N_2^\nu \equiv 0
\vspace{0.1 cm} \\
-m_g\frac{d^2 }{d\tau^2} \left( \frac{\kappa_2 \kappa_3}{\kappa_1} \right)
+ m_g \frac{\kappa_2\kappa_3^3}{\kappa_1}
-q F_{\mu \nu} T^\mu N_3^\nu \equiv 0
\end{array}
\end{equation}
\end{itemize}

For solving the equations of motion of any of
these models, one must get the trajectories of the particles in
Minkowski space by integrating all the FS equations
coupled with the geometric term of the model.

\section{Solution for a free relativistic particle: $m_g=0$.}
\label{casomg0}

Let us first consider the simplest possible case, to test the formalism,
when no geometrical Lagrangian is included. Then,
the set of Eqs. \ref{sec:acciones} reduce to

\begin{equation}
\label{mg0}
\begin{array}{c}
-m\kappa_1 - q F_{\mu \nu} T^{\mu} N_1^\nu=0\\
q F_{\mu \nu} T^{\mu} N_2^\nu=0 \\
q F_{\mu \nu} T^{\mu} N_3^\nu=0
\end{array}
\end{equation}
\noindent Let us show now that the first of this equations is the
Newton's second law with the Lorentz force in the right side. To see this, substitute
$\kappa_1$ from the first FS $\kappa_1=\eta(N_1,\frac{dT}{d\tau})$ we get

\begin{equation}
\label{eqnewton1ren}
-m\eta_{\mu \nu}N_1^\nu\, \frac{dT^\mu}{d\tau}- q F_{\mu \nu} T^{\mu} N_1^\nu= 0
\end{equation}
\noindent factorizing $\eta_{\mu \nu} N_1^\nu$ and re-arranging repeated
indices, we obtain

\begin{equation}
\label{eqnewton2ren}
\eta_{\nu \alpha} N_1^\nu \left( -m\frac{dT^\alpha}{d\tau}+
q F^\alpha_\beta T^\beta\right)  =0
\end{equation}
\noindent as $N_1^\mu$ is still an arbitrary vector, one must realize that
the expression in parentheses is always zero; and this happens because the action
Eq.~\ref{acciongeneralgeo} reduces to the usual relativistic action
if $m_g=0$, that is

\begin{equation}
\label{eqnewton}
\frac{dT^\mu}{d\tau}= \frac{q}{m} F^\mu_\nu T^\nu
\end{equation}
\noindent We are left with the second and third relations in Eqs.~\ref{mg0}. It
should be noted that for all electromagnetic fields in special
relativity, one must have

\begin{equation}
\label{cmg0}
\begin{array}{c}
F_{\mu \nu} T^{\mu} N_2^\nu=0 \\
F_{\mu \nu} T^{\mu} N_3^\nu=0
\end{array}
\end{equation}
\subsection{Constant electromagnetic field}
\label{fuvcte}

Let us consider now the case in which $F_{\mu \nu}$ is constant.
We show now from the Frenet-Serret equations ~\ref{eq:fs} that
the solution curve will have all its curvatures constants.  Let us then
define $M^\mu=\frac{dT^\mu}{d\tau}$; then according to Eq.\ref{eqnewton},
$M^\mu=q/m \, F^\mu_\nu T^\nu$. The derivative is

\begin{equation}
\label{idenMM}
\frac{dM^\mu}{d\tau}=\frac{q}{m} \, F^\mu_\nu \frac{dT^\mu}{d\tau}
=\frac{q}{m} \, F^\mu_\nu \, M^\mu.
\end{equation}
\noindent Let us take now the derivative of the magnitude of $M$, that is

\begin{equation}
\label{deridenMM}
\begin{array}{ll}
\frac{d}{d \tau} \left( \eta_{\mu \nu} M^\mu \, M^\nu \right) & =
2 \eta_{\mu \nu} M^\mu \frac{d M\nu }{d\tau} \\
&= 2 \eta_{\mu \nu} M^\mu \frac{q}{m} F^\nu_\alpha M^\alpha \\
&= 2 \frac{q}{m} \left( F_{\mu \nu} M^\mu M^\nu \right) \equiv 0
\end{array}
\end{equation}
\noindent where the third line vanishes because $F$ is antisymmetric and
constant; we may then conclude that

\begin{equation}
\label{cteMM}
\eta \left( M,M \right)\equiv
\eta \left( \frac{dT}{d\tau},\frac{dT}{d\tau} \right)=cte
\end{equation}
\noindent In what follows, we will use this result to show that
all the curvatures are constant. Let us then substitute the right hand side of
Eq.~\ref{cteMM} with the first Frenet-Serret relation

\begin{equation}
\label{cteMMkappa1}
\eta \left( \kappa_1 N_1, \kappa_1 N_1 \right)=cte=> \kappa_1^2=cte \;,
\end{equation}
\noindent which directly implies that the first curvature is
constant. Now, it is possible to
rewrite the Newton's second law, Eq.~\ref{eqnewton}, in the form

\begin{equation}
\label{eqnewtonkappa}
\kappa_1 N_1^\mu= \frac{q}{m} F^\mu_\nu T^\nu \;.
\end{equation}
\noindent Squarring it we obtain

\begin{equation}
\label{kappa1cte}
\kappa_1^2= \left( \frac{q}{m} \right)^2 \eta_{\mu \nu}
F^\mu_\alpha F^\nu_\beta T^\alpha T^\beta \equiv cte \;.
\end{equation}
\noindent We then recognize $\kappa_1$ as an invariant
whenever $F_{\mu \nu}$ is constant.  Taking the derivative of the
Eq.~\ref{eqnewtonkappa} with constant $\kappa_1 $ we learn
that the $N_1$ vector satisfies an equation of the
form ~\ref{cteMM},

\begin{equation}
\label{eqMN1}
\frac{dN^\mu_1}{d\tau}= \frac{q}{m} F^\mu_\nu N^\nu_1\;.
\end{equation}
\noindent Let us now use the second FS Eq.~\ref{fs2} and square it

\begin{equation}
\label{cteMMkappa2}
\eta \left( \frac{dN_1}{d\tau},\frac{dN_1}{d\tau} \right)
= \eta \left( \kappa_2 N_2+\kappa_1 T, \kappa_2 N_2+\kappa_1 T\right)=
\kappa_2^2-\kappa_1^2=cte
\end{equation}
\noindent from which we get that $\kappa_2$ must also be a constant. On the
other hand, by using Eq.~\ref{eqMN1} into the left hand side of Eq.~\ref{cteMMkappa2}, we
obtain the relation

\begin{equation}
\label{kappa2cte}
\kappa_2^2= \kappa_1^2+ \left( \frac{q}{m} \right)^2 \eta_{\mu \nu}
F^\mu_\alpha F^\nu_\beta N_1^\alpha N_1^\beta \equiv cte
\end{equation}

We continue the process by obtaining the term with $\kappa_2$ from the
second FS relation

\begin{equation}
\label{kappa2des}
\kappa_2 N_2^\mu=\frac{d N_1^\mu}{d \tau}-\kappa_1 T^\mu
\end{equation}
\noindent Substituting Eq.~\ref{eqMN1} on the right hand side of
Eq.~\ref{kappa2des} and taking the $\tau$ derivative on the resulting expression,
we obtain

\begin{equation}
\label{eqMN2}
\frac{dN^\mu_2}{d\tau}= \frac{q}{m} F^\mu_\nu N^\nu_2\;.
\end{equation}
\noindent Thus, the square of the derivative is a constant, in such a way that

\begin{equation}
\label{cteMMkappa2p}
\begin{array}{l}
\eta_{\mu \nu} \left( \frac{dN^\mu_2}{d\tau} \frac{dN^\mu_2}{d\tau} \right)
=cte\\
\eta_{\mu \nu} \left( \kappa_3 N_3^\mu-\kappa_2 N_1^\mu \right)
\left(\kappa_3 N_3^\mu-\kappa_2 N_1^\mu \right)=cte\\
\kappa_3^2+\kappa_2^2=cte.
\end{array}
\end{equation}
\noindent from which we conclude that $\kappa_2$ is also a constant.

Let us now consider the term with $N_3$ from the third FS relation ~\ref{fs3}

\begin{equation}
\label{eqMN3}
\frac{dN^\mu_3}{d\tau}= \frac{q}{m} F^\mu_\nu N^\nu_3
\end{equation}
\noindent by applying a similar procedure we obtain

\begin{equation}
\label{kappa3cte}
\kappa_3^2= \left( \frac{q}{m} \right)^2 \eta_{\mu \nu}
F^\mu_\alpha F^\nu_\beta N_3^\alpha N_3^\beta \equiv cte
\end{equation}
\noindent from which we conclude that $\kappa_3$ is a constant.

As we have seen, for an $F$ constant the three curvatures are constants and can be
expressed in terms of $F$ itself. In order to show the form of the
curvatures in terms of the
electric and magnetic fields, $\vec{E}$ and $\vec{B}$, respectively,  we use
the conventions given by Weinberg\cite{wein} for the tensor $F$, see the
Appendix ~\ref{app}.

For any vector $A^\mu$, the contraction with the
electromagnetic tensor, in the form suggested by the right
hand side of the curvatures is given by

\begin{equation}
\label{ffaa}
\eta^{\alpha \beta} F_{\alpha \mu} F_{\beta \nu} A^\mu A^\nu=
-\left( \vec{A} \cdot \vec{E} \right)^2+\left( A^t \vec{E}
+ \vec{A} \times \vec{B} \right)^2
\end{equation}
\noindent Then in the coordinate system attached to the laboratory, the tangent vector
is

\begin{equation}
\label{tangente}
T^\mu=\left(\gamma,\gamma \vec{v} \right)
\end{equation}
\noindent where $\vec{v}$ is the particle velocity and

\begin{equation}
\label{gamaexpl}
\gamma=\frac{1}{\sqrt{1-\vec{v}^2}}.
\end{equation}
\noindent Therefore, the invariant for the fist curvature is

\begin{equation}
\label{invkappa1}
\kappa_1^2= \left( \frac{q \gamma }{m} \right)^2
\left[ -\left( \vec{v}\cdot \vec{E}\right)^2 + \left( \vec{E} + \vec{v} \times \vec{B} \right)^2 \right]
\end{equation}
\noindent It is easy to complete the dot product in the first term of the
right hand side, then Eq.~\ref{invkappa1} can be written in the form

\begin{equation}
\label{invkappa1lor}
\kappa_1^2= \left( \frac{\gamma }{m} \right)^2
\left[ \vec{F}^2 - \left( \vec{v} \cdot \vec{F} \right)^2 \right]
\end{equation}
\noindent where $\vec{F}$ is the Lorentz force
$\vec{F}=q\left(\vec{E} - \vec{v} \times \vec{B} \right)$. Equation ~\ref{invkappa1lor}
was obtained by \cite{honig} and \cite{lopez}.

\section{Conclusions}
\label{sec:conclu}

In this paper we have obtained the equations of motion for a charged
particle moving in an electromagnetic field when its action includes
terms with its own world-line curvatures. This dynamical problem is
mathematically difficult even on how to write the equations of motion,
particularly when they are in terms of the particle embedding functions
$X(\tau)$. An alternative approach useful for handling the
mathematics of this problem is based on taking advantage of the
Frenet-Serret frame, as we have shown here. We have obtained the
Frenet-Serret equations in Minkowski space-time and then we have used them
in order to develop a variational calculus well adapted for
tackling this kind of problems.

The problem of a relativistic particle moving in an electromagnetic
field is interesting both in theoretical and applied physics. For
instance, in plasma physics, this problem is concerned with the
mechanics of particle acceleration, when heating and radioactive
effects are taken into account, see Ref.\cite{ondarza}. This kind of
studies are based on the numerical integration of Eq.~\ref{eqnewton}
obtained in Section~\ref{casomg0}, where we also proved the
equivalence of the alternative approach. It would be interesting to
compare the results obtained by integrating the two formalisms.

Finally, we mention that all the mathematical formalism
developed in Sects. \ref{subsec:tFSb}
and \ref{subsec:pertur} can easily be translated into
an Euclidean space, where the curvature dependent actions
and the world-line of the relativistic particle would be replaced by
curvature dependent energy functionals and by smoothly continuous curves,
respectively, see Refs.~\cite{kamien} and references there in.

The space of solutions in the Euclidean frame is abundant
and physically interesting, as can be seen
in Ref.~\cite{elastica}, where the equilibrium configurations of a $2D$ closed
rigid loop were studied. In Ref.~\cite{hamil} the integrability of
some curvature dependent energy functionals was established
by making use of the constants of integration obtained by applying the
Noether's theorem. Subsequently, in Ref.~\cite{showcase}, the equilibrium
configurations curves in the Euclidean $3D$ space were numerically obtained for
the models of Ref.~\cite{hamil}.  
\appendix
\section{Appendix}
\label{app}

For the electromagnetic tensor, we follow the notation
of Weinberg \cite{wein}, we then have

\begin{equation}
\label{matrizFgen}
F_{\mu \nu}=
\left(
\begin{array}{cccc}
0& -E_1& -E_2& -E_3\\
E_1& 0 & B_3& -B_2\\
E_2& -B_3& 0 & B_1\\
E_3& B_2& -B_1& 0
\end{array}
\right)
\end{equation}
\noindent and its dual ${}^*F_{\mu \nu}=\frac{1}{2} \epsilon_{\mu
\nu \alpha \beta} \, F^{\alpha \beta}$ and in matrix notation is:

\begin{equation}
\label{matrizFmat}
{}^*F_{\mu \nu}=
\left(
\begin{array}{cccc}
0& -B_1& -B_2& -B_3\\
B_1& 0 & -E_3& E_2\\
B_2& E_3& 0 & -E_1\\
B_3& -E_2& E_1& 0
\end{array}
\right)
\end{equation}
\noindent The field invariant are then the following

\begin{equation}
\label{invaF}
\begin{array}{c}
F_{\mu \nu}\, F^{\mu \nu}= -2 \left(\vec{E}^2-\vec{B}^2 \right) \\
{}^*F_{\mu \nu} F^{\mu \nu}= -2 \vec{E} \cdot \vec{B}
\end{array}
\end{equation}
\section{Acknowledgments}

GA thanks to UNISON for financial support during the performance of this project.


\begin{thebibliography}{99}


\bibitem{landau} Landau, L.D. and Lifshitz, E.M., {\it A shorter course of theoretical
Physics}, Vol. 1, Mechanics and Electrodynamics. Pergamon Press, 1975.

\bibitem{wein} Weinberg, S.,{\it  Gravitation and cosmology: principles
and applications of the general theory of relativity}, John Wiley and Sons, 1972.

\bibitem{polyakov} Polyakov,A. {\it Nucl. Phys.}, {\bf B268}, (1986), 406.


\bibitem{fsdyn} Arreaga G., Capovilla R., Guven J., {\it Class. Quant. Grav.}, {\bf 18}, (2001), 5065-5083.

\bibitem{polyakovl} Polyakov, A.M., {\it Gauge Fields and Strings.}, (1987), New York:Harwood Academic.

\bibitem{canonical} Plyushchay, M.S., {\it Mod.Phys.Lett.}, {\bf A 3}, (1988), No.13, 1299-1308.


\bibitem{massive} Plyushchay, M.S., {\it Int.Jour.Mod.Phys.}, {\bf  A 4}, (1989), No.15, 3851-3865.


\bibitem{dereli} Dereli,T., Hartley, D.H., Onder, M. and Tucker, R.W., {\it Phys.Lett.}, {\bf B 252}, (1990), 601-.

\bibitem{neste} Nesterenko, V.V. , {\it J.Phys.}, {\bf A 22}, (1989), 1673-.
Nesterenko, V.V. , {\it J.Math.Phys.}, {\bf 32}, (1991), 3315-.
Nesterenko, V.V. , {\it Int.Jour.Mod.Phys.}, {\bf A 6}, (1991), 3989-.
Nesterenko, V.V. , {\it Phys.Lett.}, {\bf B 327}, (1994), 50-.

\bibitem{neste2} Nesterenko, V.V., Feoli, A. and Scarpetta, G., {\it Class.Quant.Grav.},
{\bf 13}, (1996), 1201-.

\bibitem{massless} Plyushchay, M.S., {\it Mod.Phys.Lett.}, {\bf  A 4}, (1989), No.9, 837-847.

\bibitem{zoller} Zoller, D. , {\it Phys.Rev.Lett.}, {\bf 65 }, (1990), 2236-.

\bibitem{quanti} Plyushchay, M.S., {\it Phys.Lett.}, {\bf  B253 }, (1991), No.1,2, 50-55.

\bibitem{otro} Plyushchay, M.S., {\it Phys.Lett.}, {\bf  B243 }, (1990), 383-388.

\bibitem{kuzne} Kuznetsov, Y. A. and Plyushchay, M.S., {\it Nucl.Phys.}, {\bf  B389}, (1993), 181.

\bibitem{comment} Plyushchay, M.S., {\it hep-th/9810101.}, (1998).

\bibitem{curvyt1} Kuznetsov, Y. A. and Plyushchay, M.S., {\it Phys.Lett.}, {\bf  B297}, (1992), 49-54.

\bibitem{curvyt2} Kuznetsov, Y. A. and Plyushchay, M.S., {\it J.Math.Phys}, {\bf  35}, (1994), 2772-2784.

\bibitem{majorana} Plyushchay, M.S., {\it Elect.J.Theor.Phys.}, {\bf  3N 10}, (2006), 17-31.

\bibitem{partorsion1} Plyushchay, M.S., {\it Nucl.Phys.}, {\bf  B362}, (1991), 54-72.

\bibitem{partorsion2} Plyushchay, M.S., {\it Phys.Lett.}, {\bf  B262}, (1991), 71-78.

\bibitem{mono} Plyushchay, M.S., {\it Nucl.Phys.}, {\bf  B589}, (2000), 413-439.

\bibitem{iden} Plyushchay, M.S., {\it Mod.Phys.Lett.}, {\bf  A 10}, (1995), 1463-1469.

\bibitem{barros} Barros, M., {\it General Relativity and Gravitation}, {\bf 34}, (2002), 837-853.

\bibitem{angel} Fernandez, A., Gimenez, A. and Lucas, P., {\it Phys.Lett.}, {\bf B543}, (2002), 311-317.

\bibitem{josu} Arroyo, J., Barros, M. and Garay, O., {\it Gen.Rel.Grav.}, {\bf 36}, (2004), 1441-1451.

\bibitem{ferra} Ferrandez, A., Guerrero, J., Javaloyes, M.A. and Lucas, P., {\it J.Geom.Phys.}, {\bf 56}, (2006), 1666-1687.

\bibitem{lopez} Lopez-Bonilla, J.L. and Pi\~na-Garza, E., {\it Part\'{\i}culas
Cl\'asicas Cargadas en Relatividad Especial}, Direccion de Publicaciones
del IPN, Mexico, 1980.

\bibitem{fomiga} Fomiga, J.B. and Romero, C., {\it Am.Journ.Phys},
{\bf 74}, (2006), 1012-1016.

\bibitem{eisen} Eisenhart L.P.,{\it An Introduction to
Differential Geometry}, Princeton University Press, 1947.

\bibitem{conslaws} Arreaga G., Capovilla R. and  Guven J., {\it Annals of Physics.},
{\bf 279}, (2000), 126-158.

\bibitem{roch} Rohrlich, F., {\it Classical charged particles}, World
Scientific Pub. Co. Inc., 2007.

\bibitem{intri} Ringermacher, H.I., {\it Physics Letters}, {\bf 74\, A}, (1979), 381-383.

\bibitem{honig} Honig, E., Schucking, E., Vishveshwara, C., {\it Journ. Math.
Physics}, {\bf 15}, (1974), 774.

\bibitem{ondarza} Ondarza R. and Gomez, F. , {\it IEEE Transactions on
Plasma Science}, {\bf 32}, Num.2, (2004), 808.

\bibitem{kamien} Kamien R, {\it Rev. Mod. Phys.}, {\bf 74},(2002),953-971.

\bibitem{elastica} Arreaga G., Capovilla R., Chryssomalakos C., and
Guven J., {\it Phys. Rev.}, {\bf E 65}, (2002), 031801.

\bibitem{hamil} Capovilla R., Chryssomalakos C., and Guven J.,
{\it Journ. Phys.}, {\bf A 35}, (2002), 6571-6587.

\bibitem{showcase} Arreaga-Garcia, G., Villegas-Brena, H. and Saucedo-Morales, J.
{\it J.Phys.A: Math.Gen.}, {\bf 37}, (2004), 1-20.

\end{thebibliography}
\end{document}